\documentclass[%
reprint,
 amsmath,amssymb,
 aps,
showkeys,
]{revtex4-2}

\usepackage{graphicx}
\usepackage{dcolumn}
\usepackage{bm}
\usepackage{siunitx}
\usepackage{hyperref}

\begin{document}

\preprint{APS/123-QED}

\title{Compact Rare-Earth Superconducting Cyclotron}
\thanks{This work was funded by the UK Science and Technology Facilities Council under Grant Nos. ST/G008248/1 and ST/R002142/1.}%

\author{Jacob Kelly}
\email{jacob.kelly@postgrad.manchester.ac.uk}
\author{Hywel Owen}
\affiliation{School of Physics and Astronomy, University of Manchester, M13 9PL, United Kingdom}
\affiliation{Cockcroft Institute of Science and Technology, WA4 4AD, United Kingdom}

\author{Timothy A. Antaya}
\author{Chris Jones}
\author{Paul Ruggiero}
\affiliation{Antaya Science and Technology, Hampton, NH 03842}


\date{\today}

\begin{abstract}
The use of superconductivity is well-known as a method to increase the average
field of a cyclotron and thereby to allow a substantial reduction of its size and mass.
We present a compact high-field design for the first superconducting cyclotron with rare-earth (holmium)
poles. Our design supports stable acceleration of protons to
$E_\textrm{k}=$ \SI{70}{\MeV} with no significant limit in beam current, 
suitable for wide applications in ocular therapy, isotope production,
radiobiological studies and nuclear physics.

\end{abstract}


\maketitle

\section{Introduction}
The cyclotron -- invented nearly ninety years ago by Livingston and Lawrence \cite{Lawrence1930, Lawrence1931, Lawrence1936} -- remains the workhorse apparatus for delivering protons and ions at moderate kinetic energies. Their twin advantages are \textit{robust simplicity} and the ready capability for \textit{high intensity}; once manufactured, a single energising coil and ion source may deliver a reliable high current of particles, often at a single extracted energy. A key advance in cyclotron technology has been the steady adoption and improvement of superconducting technology, the first superconducting cyclotron being realised in the K500 by Blosser et al. \cite{Blosser1989}. In recent years several superconducting cyclotrons extracting protons with kinetic energies up to $E_\textrm{k}$ = \SI{250}{\MeV} have been developed with the aim of making proton therapy systems more compact and affordable~\cite{JKim2001, Schillo2001, Schippers2004, Roecken2010}; this justification for using superconductivity in particle therapy has been established in a number of reviews~\cite{Friesel2009, Alonso2012, Krischel2012, Owen2014, Owen2014a, Owen2016}. Indeed, the highest dipole field obtained in a particle accelerator of any type today is the \SI{9}{\tesla} achieved in the Mevion medical synchrocyclotron \cite{Smirnov2016}, of which there are several commercial examples. Superconducting cyclotrons have also been developed for lower extraction energies -- particularly for isotope production~\cite{Smirnov2014} -- and for ion therapy \cite{JKim2001a, Calabretta2006, Jongen2010, HWKim2016}. An advantage of high field in a low-energy cyclotron (say, at 12~MeV) is that the complete magnet (including yoke) may be placed within a compact cryostat~\cite{Smirnov2014}.

At moderate to high energies the challenge remains of how to simultaneously obtain both a high average field $B$ -- which allows the overall mass and volume of the cyclotron to reduce (roughly as $1/B^3$) -- whilst also creating a suitable field profile and focusing to give isochronous behaviour and thereby to allow for the highest proton intensities~\cite{Krischel2012, Smirnov2016}. This is the subject of the present paper, in which we discuss a novel method of creating an isochronous, high-field cyclotron based on a rare-earth `flying pole'; we study an example isochronous \SI{70}{\MeV} design that utilises a \SI{4.52}{\tesla} central field. We show the first realistic method of combining a cold holmium pole (which enables the strong focusing at high fields) with a superconducting NbTi coil and warm yoke. The result delivers high currents of \SI{70}{\MeV} protons from a cyclotron of unprecedentedly small size; such a design enables low-energy proton therapy -- for example ocular therapy~\cite{Kacperek2012} or surface lesions -- at high dose rates that would also enable techniques such as FLASH therapy. Intensities of several hundred microamperes would also allow it to be used for the generation of medical isotopes. A source of \SI{70}{\MeV} protons is also very attractive for uses in radiobiological research, as it provides a sample penetration depth well-suited for typical experimental geometries.

\subsection{Early Cyclotrons}
The development of the cyclotron in 1930 was inspired by the early nuclear structure research of Rutherford and Chadwick, where various groups looked at how to increase bombarding particle energies beyond what could be achieved with static voltages.  Using a magnetic field to resonantly circulate protons through a moderate radio-frequency (RF) voltage of \SI{4}{\kV}, Lawrence and Livingston achieved acceleration to \SI{1.22}{\MeV}, comparable with direct voltage methods of the time.

Theoretical advances allowed Ernest Lawrence's group at Berkeley to construct cyclotrons of steadily-increasing energy and current. Early work addressed the loss in resonance of the accelerated particles with the RF in a uniform magnetic field, due to their relativistic mass increase by the factor $\gamma=E/m_0 c^2$; the increased mass $\gamma m_0$ decreases the angular frequency $\omega=qB_0/\gamma m_0$ of a particle of charge $q$ and rest mass $m_0$ as it obtains total energy $E$ in a magnetic field $B_0$ \cite{Bethe1937}. Two solutions were successful. The first proposal was the isochronous cyclotron (Thomas, 1938 \cite{Thomas1938}) in which the magnetic field $B$ is scaled with radius as $B=\gamma B_0$, allowing the RF to operate at a constant frequency $\omega_0$. It was determined that whilst an azimuthally-symmetric isochronous design provides phase stability it does not give axial (vertical) stability of ions. Thomas showed that in an isochronous cyclotron an azimuthally-varying-field (AVF) is necessary for that axial stability; this AVF was obtained by shaping the cyclotron pole into several sectors of hills (higher field) and valleys (lower field). Kerst later showed in 1956~\cite{Kerst1956} that introducing spiral sectors to an AVF design can increase the overall axial focusing. 

The second proposal was the synchrocyclotron (Bohm and Foldy, 1947~\cite{Bohm1947}) in which the RF frequency $\omega$ is modulated as $\omega=\omega_0/\gamma$ while a single bunch of particles is accelerated, limiting the output intensity but simplifying the magnet; in a synchrocyclotron $B$ does also decrease somewhat with radius to give the required condition for weak focusing of the accelerating particles. Later developments during the 20th century saw isochronous cyclotrons and synchrocyclotrons accelerate charged particles to energies up to \SI{1.2}{\GeV} and with currents up to several milliamperes; however, higher-energy cyclotrons were initially large because of the limited magnetic fields available then.

\subsection{Superconducting Cyclotrons}
In the mid-1970s groups at both the National Superconducting Cyclotron Laboratory at Michigan State University (MSU) and the Chalk River Nuclear Laboratory began work on and construction of the first superconducting isochronous cyclotrons. These were the K500 (first operated in 1982 at MSU) and the K520 at Chalk River (first operated in 1985) \cite{Blosser1975}. The K1200 isochronous cyclotron weighing \SI{280}{tonnes} followed at MSU in 1984 with a bending power greater than that of the lower-field normal-conducting \SI{7800}{tonnes} synchrocyclotron at Gatchina.

It is notable that cyclotrons have always been intimately linked with the field of medicine. John Lawrence (the brother of Ernest) joined the Berkeley group in 1935 and pioneered methods of cancer treatment, both by the application of cyclotron-produced short-lived radioisotopes, and by the bombardment of tumours with protons and neutrons \cite{JHLawrence1937}. Fifty years later the superconducting cyclotron group at MSU began work on the K100 compact superconducting cyclotron to accelerate deuterons to \SI{50}{\MeV} and onto a beryllium target for neutron therapy. This device was installed at Harper Hospital (Detroit) and was the first physically-rotating (gantry-mounted) cyclotron \cite{Blosser1989}. These early designs have inspired the present-day commercial superconducting cyclotrons used for particle therapy that are exemplified by the Varian COMET (isochronous, \SI{2.4}{\tesla}, \SI{250}{\MeV}), IBA S2C2 (synchrocyclotron \SI{5.7}{\tesla}, \SI{230}{\MeV}), and Mevion (synchrocyclotron \SI{9}{\tesla}, \SI{250}{\MeV}).

When considering a superconducting cyclotron at \SI{70}{\MeV}, it is useful to compare it to existing normal-conducting systems. We use the example of the Scanditronix MC-62, a \SI{62}{\MeV} normal-conducting cyclotron presently used for ocular therapy at the Clatterbridge Centre for Oncology, UK. This cyclotron has an outer yoke diameter of \SI{4}{\m} and a magnet mass of 120~tons, typical for a central field of \SI{1.8}{\tesla}~\cite{scanditronixmanual}. In comparison, the 70~MeV compact design discussed in this paper is projected to have a diameter of \SI{1.34}{\m} and a mass of less than \SI{9}{tonnes}. The following calculation is a more general demonstration of the drastic reduction in size of a cyclotron at a given energy as a result of the higher $B$-field possible with superconducting coils.

\begin{figure*}[!htb]
    \centering
    \includegraphics{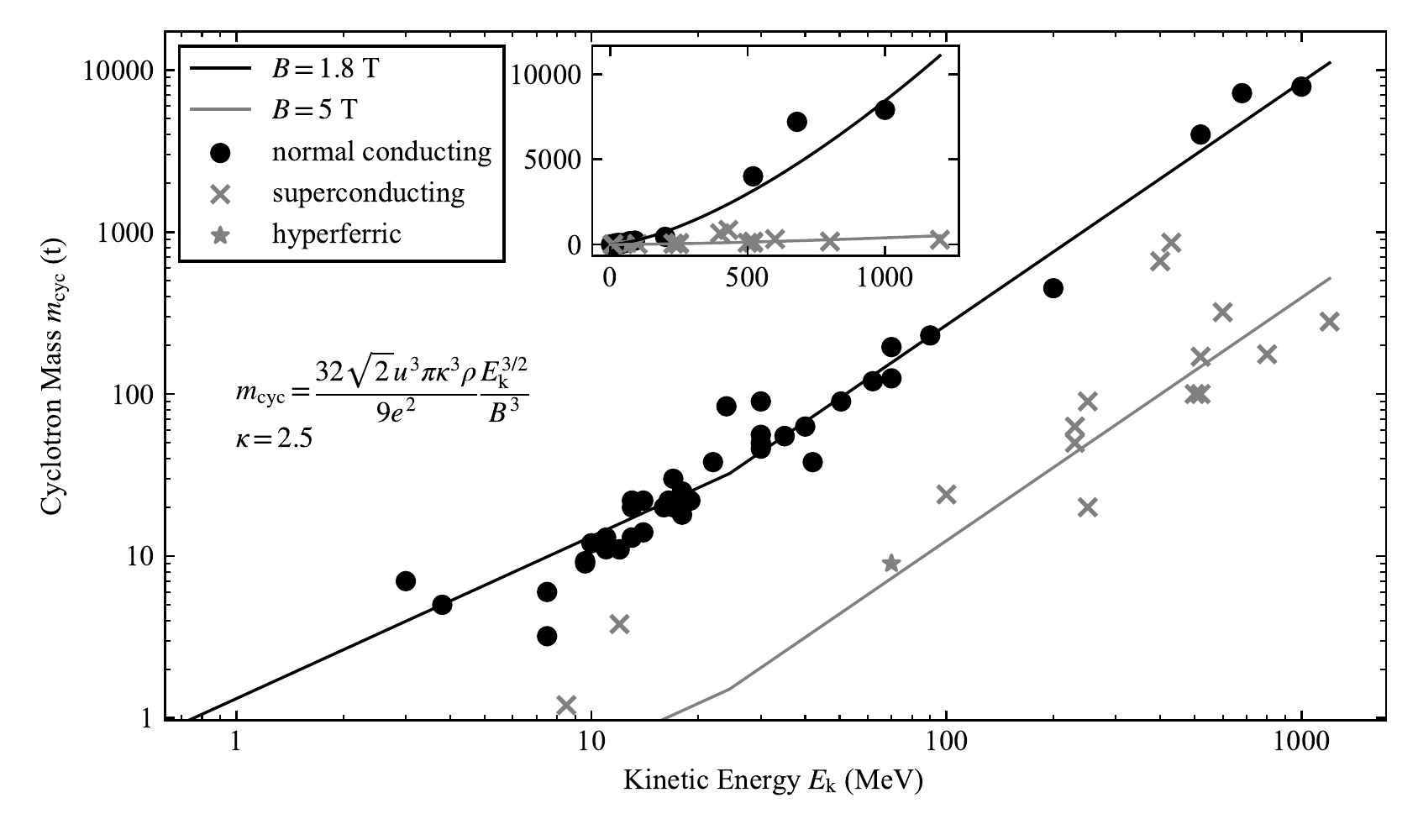}
    \caption{Logarithmic and linear (inset) plots of a survey of cyclotron masses, as a function of extracted proton energy $E_\mathrm{k}=K(Q^2/A)$, are shown with an approximate expression which predicts the mass of a cyclotron for a given energy and $B$-field. High-field superconducting cyclotrons are seen to be significantly more compact than normal-conducting cyclotrons. We also show our design (`hyperferric') which is discussed in detail in the text.}
    \label{fig:cyclotron_masses}
\end{figure*}

The kinetic energy $E_\textrm{k}$ of an ion with mass $A$ (in a.m.u.) and charge $Q$ (in units of $e$) at extraction radius $r_\mathrm{ext}$ in a cyclotron of field strength $B$ is given by
\begin{equation}
    E_\textrm{k}=\frac{(eBr_\mathrm{ext})^2}{2u}\left(\frac{Q^2}{A}\right)=K\left(\frac{Q^2}{A}\right)\textrm{,}
    \label{eq:energy_k}
\end{equation}
where $u$ is 1~a.m.u. Given Eq.~\ref{eq:energy_k} we may construct an approximate expression for the mass of a cyclotron that accelerates protons ($Q^2/A=1$) to a kinetic energy $E_\mathrm{k}$. We assume that the cyclotron's steel yoke (of density $\rho$) is spherical with an outer radius $r_\mathrm{cyc}$, which is related to the particle extraction radius by a factor $\kappa=r_\mathrm{cyc}/r_\mathrm{ext}$. The mass of the sphere $m_\mathrm{cyc}=4\pi r_\mathrm{cyc}^3 \rho/3$ is related to $E_\mathrm{k}$ and $B$ as
\begin{equation}
    m_\mathrm{cyc}=\frac{32\sqrt{2}u^3\pi \rho \kappa^3}{9e^2}\frac{E_\textrm{k}^{3/2}}{B^3}\textrm{,}
    \label{eq:energy_mass}
\end{equation}
which shows the $1/B^3$ scaling at a given $E_\textrm{k}$. At extraction, cyclotrons with resistive coils are limited practically to a magnetic flux density $B<$ \SI{2}{\tesla}. Superconducting cyclotrons offer a much higher flux density at extraction, often as much as $B=$ \SI{5}{\tesla}, and in the case of the Mevion S250 synchrocyclotron have achieved fields over $B=$ \SI{9}{\tesla}. Figure~\ref{fig:cyclotron_masses} compares the derived relation of Eq.~\ref{eq:energy_mass} to data obtained from a representative range of research and commercial, normal conducting and superconducting cyclotrons. The predicted position on this plot of the \SI{70}{\MeV} design discussed in this paper is shown. We see that high $B$-field superconducting magnets can obtain a cyclotron of a given $E_\mathrm{k}$ that is at least \emph{an order of magnitude lighter} than the corresponding normal-conducting equivalent.  

Low-energy, isochronous, superconducting cyclotrons ($E_\textrm{k}<$~\SI{20}{\MeV}) may rely on the AVF produced by an ordinary steel pole to provide sufficient axial focusing of ions, whereas higher-energy isochronous cyclotron designs (with energies perhaps larger than $E_\textrm{k}=$~\SI{100}{\MeV}) require additional axial focusing which can be provided by so-called flutter coils that increase the AVF. In the intermediate energy range the use of a high-permeability material for the cyclotron pole is proposed here as a means of generating the necessary AVF. Rare-earth metals such as gadolinium and holmium are candidate pole materials, each having saturation magnetisations significantly higher than that of low-carbon steel; gadolinium has the advantage of having a much higher Curie temperature but holmium saturates at a higher field. 

\subsection{Rare-Earth Superconducting Magnets}
\begin{figure}[!htb]
\centering
\includegraphics[width=\columnwidth]{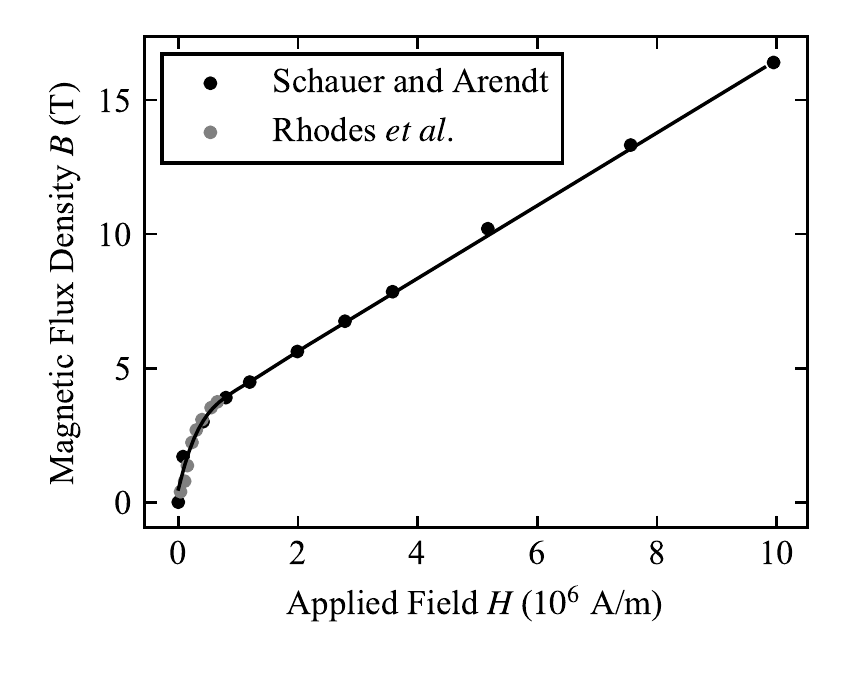}
\caption[]{Holmium $B$-$H$ curve at \SI{4.2}{\kelvin}; data taken from Schauer and Arendt \cite{Schauer1983} and Rhodes \textit{et al.} \cite{Rhodes1958}, fit from Norsworthy \cite{Norsworthy2010}. The onset of saturation at 3.9~T may be seen.}
\label{fig:holmiumbh}
\end{figure}

Holmium is a rare-earth metal which undergoes an anti-ferromagnetic to ferromagnetic phase transition with decreasing temperature at around approximately \SI{20}{\kelvin}. In the ferromagnetic state, and at \SI{4.2}{\kelvin}, holmium has the highest saturation magnetisation of any element: $\mu_0M_\mathrm{s}=$ \SI{3.9}{\tesla}. A 1958 paper by Rhodes, Spedding, and Levgold characterised the magnetic dependence of holmium on temperature (down to \SI{4.2}{\kelvin}) and on applied magnetic field (up to $\mu_0H=$ \SI{1.6}{\tesla}) \cite{Rhodes1958}; these magnetic measurements were performed on a torus of rectangular cross-section around which a normal conductor was wound and a current applied \cite{Levgold1953}. A 1983 paper by Schauer and Arendt then characterised the $B$-$H$ curve of holmium at \SI{4.2}{\kelvin} in a much stronger applied magnetic field (up to $\mu_0H=$ \SI{12.5}{\tesla}) \cite{Schauer1983}; in this case, two holmium cylinders were placed as flux concentrators within a Nb$_3$Sn superconducting solenoid. A gap between the cylinders of \SI{5.5}{\mm} allowed field measurements to be taken using a Hall probe. The $B$-$H$ data from both papers are in good agreement and are shown in Fig.~\ref{fig:holmiumbh} with a fit from Norsworthy \cite{Norsworthy2010}.

There have been various implementations of rare-earth metals for field enhancement in superconducting magnets. Examples include the use of holmium both as a flux concentrator in superconducting solenoids~\cite{Schauer1983} and as a pole piece in place of traditional iron alloys \cite{Hoard1985}. In these cases rare-earth metals were used to boost already high-field systems beyond the quench limit of commercially-available superconductors, which lies around \SI{8}{\tesla} for NbTi and \SI{13}{\tesla} for Nb$_3$Sn. In contrast, in our present study we operate the superconducting coils well below the quench limit of NbTi and we use the large \SI{3.9}{\tesla} saturation magnetisation of holmium to create the required azimuthally-varying $B$-field; using only (saturated) iron poles would give insufficient AVF.

Further examples of the use of rare-earth metal poles have been in high-gradient superconducting quadrupoles for linacs~\cite{Barlow1992, Kim1995}, in a \SI{7.5}{\tesla} quadrupole design for magnetic circular dichroism experiments~\cite{Chang2002}, and in an octupole design for photon scattering experiments~\cite{Hwang2002}. Holmium has also been proposed as a material for use in superconducting wigglers for beam emittance reduction in damping rings, for example in the Compact Linear Collider in which a larger $B$-field decreases the minimum achievable emittance~\cite{Peiffer2010, Schoerling2012}. However, whilst designs exist for superconducting rare-earth cyclotron and cyclotron-like accelerators, an example has yet to be realised. One such design was a compact FFAG accelerator for \SI{400}{\MeV\per\atomicmassunit} carbon ions~\cite{Qin2011}, which proposed gadolinium poles to increase the possible field gradient whilst maintaining a near-room-temperature magnet; in this design the poles are in direct contact with the iron yoke. However, it is not clear in this design how the entire magnet could be maintained sufficiently below the Curie temperature of c.~\SI{290}{\kelvin} to make full use of the Gd poles.

Our present design differs in that it utilises both a superconducting coil and a (cold) rare-earth pole -- holmium in our design -- in conjunction with a (warm) iron yoke. We propose the term `hyperferric' for such a magnet; ordinary superferric magnets have iron-dominated fields in which the driving coils may be either resistive or superconducting, whereas we have used a rare-earth pole instead of iron. Physically separating the holmium pole from the iron yoke allows it to be cooled with a small cold mass and volume, in what we term a `flying pole' configuration; this offers enormous practical advantages for any cyclotron of \SI{70}{\MeV} or greater extraction energy, since the yoke (which constitutes the bulk of the overall cyclotron mass) can be kept at room temperature. The superconducting coil produces the large average field required for a compact overall magnet size, whilst the shaped holmium poles allow the largest field variation between the hills and valleys. Whilst the holmium poles need to be cooled significantly below their Curie temperature of \SI{20}{\kelvin} -- actually to \SI{4.2}{\kelvin} to maximise their saturation magnetisation -- this separation of pole and yoke (the flying pole) offers some advantages in the magnet design as we shall see below. The nearby superconducting coils must already be cooled, and so it is not onerous to also cool the poles; one may use a pair of cryostats either side of the room-temperature dees and beam vacuum vessel.

An earlier design -- the so-called Megatron K250 -- aimed at a proof-of-principle \SI{250}{\MeV} proton cyclotron with holmium poles~\cite{Norsworthy2010, Zhang2011}; such a cyclotron could have applications in high-dose-rate proton therapy and for detection of fissile materials. Whilst an isochronous field and sufficient flutter was obtained, this design was not yet practical and had insufficient internal aperture to accommodate the dees, cryostat and accelerated beam. The design presented here resolves this with a larger pole gap of \SI{5.2}{\cm}, and we believe it is the first practical design of a so-called hyperferric cyclotron. 

\section{Magnet Design of a Hyperferric Cyclotron at 70~\NoCaseChange{MeV}}
\label{sec:magnet_design}
We have applied the flying-pole `hyperferric' approach described above in a feasible design with the aim of constructing a prototype; we have studied a \SI{70}{\MeV} proton extraction energy since increasing the central field from around 1.8~T (normal conducting) to \SI{4.52}{\tesla} (hyperferric) drastically reduces the size of such a system. As we saw above, a \SI{4.52}{\tesla}, \SI{70}{\MeV} cyclotron is less than one-tenth the mass of its normal-conducting equivalent, and turns such a proton source from the preserve of regional facilities to a system that could readily be installed in a small laboratory; the cost often scales quite closely with the cyclotron mass. 

The use of a flying pole allows us to retain a room-temperature yoke constructed from ordinary 1010 steel whilst also allowing the use of holmium poles. Holmium is chosen over gadolinium due to its larger saturation field, and can be accommodated in the overall design since the coils are already superconducting (so that a common cryogenic system is possible), and sufficient AVF can be obtained across a realistic pole gap. The main cyclotron parameters are shown in Table~\ref{tab:cyclotronparameters}, where we note in particular the remarkably-small yoke size and mass.

Our design uses three spiral sectors to provide axial focusing. The key features of our design are:
\begin{itemize}
    \item The use of a single pair of holmium pole tips, with sufficient space for a cryostat, accelerating structure, and structural support which has not previously been described for a cyclotron;
    \item Physical separation of the holmium poles from the yoke (the `flying pole' approach), which allows the yoke to operate at room temperature whilst the holmium poles operate at \SI{4.2}{\kelvin};
    \item Shaping of the \textit{reverse} (non-orbit-facing) side of the holmium poles (which we term `back-cut poles') to adjust the field profile, increasing the volume of ferromagnetic material close to particles orbiting in the median plane and thus enabling stronger focusing;
    \item The resulting small overall cyclotron size allows each steel yoke-and-pole half to be manufactured from a single forging, eliminating some field errors that can arise from misaligned assembly of several yoke pieces.
\end{itemize}

Three spiral sectors provide axial focusing at the larger orbit radii, with a conventional dee and stem arrangement for acceleration. Protons may be axially injected from an external \SI{10}{\GHz} ECR ion source delivering approximately \SI{10}{\keV} protons through a conventional spiral inflector and puller arrangement. Injection occurs in a weak-focusing cone field approximately \SI{2}{\percent} greater than the central field value $B_0$ of \SI{4.52}{\tesla}. Conventional particle extraction using electrostatic deflectors will allow currents of at least \SI{100}{\nA} at good extraction efficiencies of perhaps \SI{80}{\percent}, sufficient for high-dose-rate particle therapy at many \SI{}{\gray\per\second}; however, self-extraction could increase the extracted current to several hundred microamperes and thereby allow use of this cyclotron for isotope production using higher-energy protons up to \SI{70}{\MeV}.

\begin{table}[!hbt]
\caption{70 MeV cyclotron parameters.}
\centering
	\begin{ruledtabular}
        \begin{tabular}{l l}
        \bf{Parameter} & \bf{Value} \\
        \hline
        Accelerated Species & Protons\\
        Extraction Energy & \SI{70}{\MeV}\\
        Extracted Current & At least \SI{100}{\nano\A} \\
        Ion Source & External ECR, \SI{10}{\GHz} \\
        Central Field & \SI{4.52}{\tesla}\\
        Pole Layout & 3-fold, Archimedean Spiral \\
        RF System & 3 dees, \SI{40}{\kV} Per Crossing \\
        Cyclotron Frequency & \SI{69}{MHz} \\
        Harmonic Number & 3\\
        Pole Gap at Hills & \SI{26}{\mm} \\
        Extraction Radius & \SI{254}{\mm}\\
        Pole Radius & \SI{310}{\mm} \\
        Yoke Radius & \SI{670}{\mm} \\
        Yoke Height & \SI{862}{\mm} \\
        Yoke Mass & $<$~\SI{9} tonnes \\
        \end{tabular}
        \end{ruledtabular}
\label{tab:cyclotronparameters}
\end{table}

Magnet modeling has been performed with \textsc{opera} using the \textsc{tosca} solver~\cite{operamanual}, both well-proven for such designs. The cyclotron magnet comprises four parts:
\begin{itemize}
    \item Low-carbon 1010 steel yoke;
    \item A pair of superconducting NbTi energising coils;
    \item Back-cut flying holmium pole pair;
    \item Central 1010 steel ring to provide weak focusing at injection.
\end{itemize}

\begin{figure*}[!htb]
\centering
\includegraphics[width=155mm]{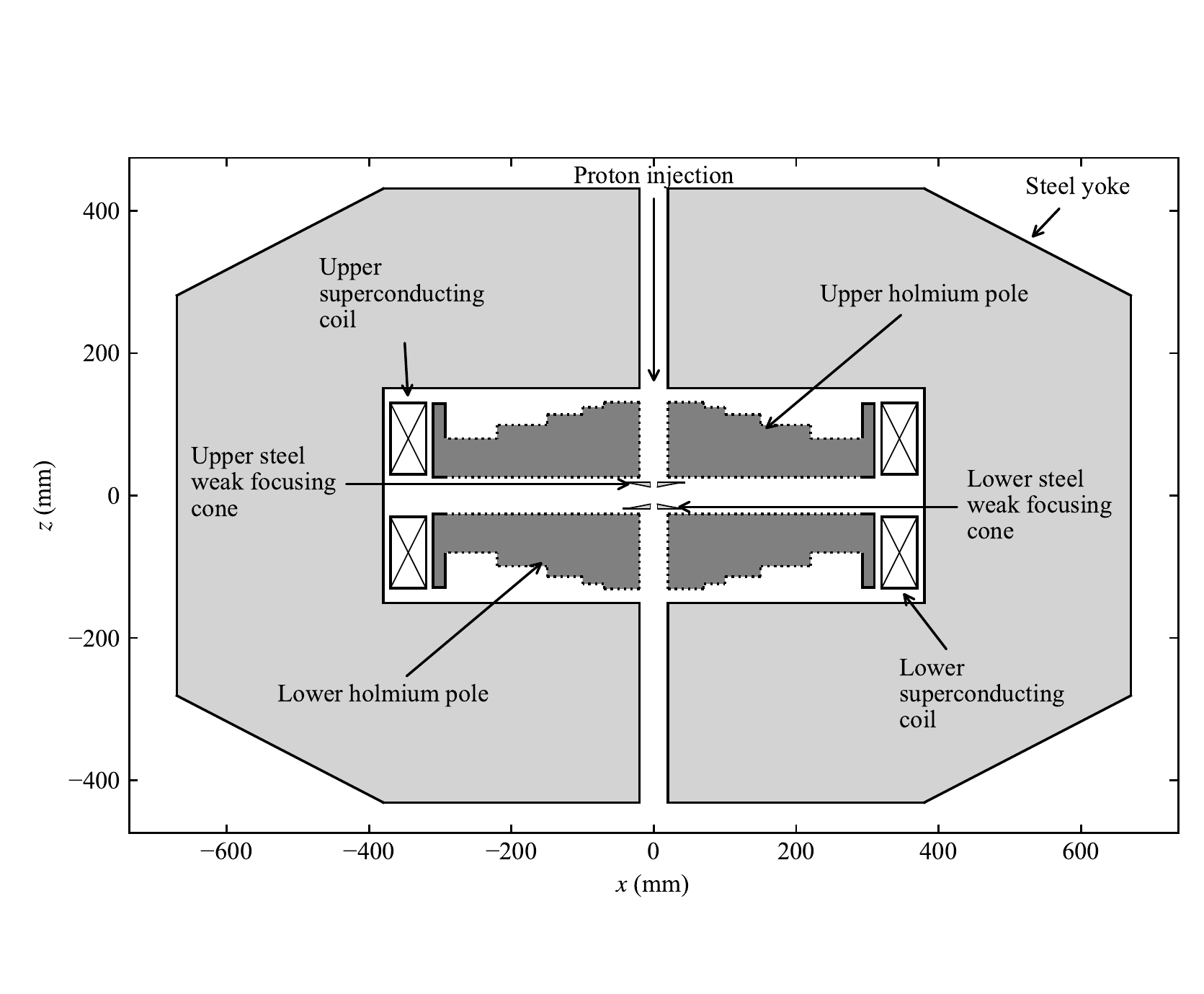}
\caption[]{Cross-sectional schematic of the magnetic materials in the \SI{70}{\MeV} proton cyclotron. The holmium poles and coils reside in a pair of cryostats operating at \SI{4.2}{\kelvin}. The yoke, coils, and steel weak-focusing cone are cylindrically-symmetric, whilst the holmium flying poles have three-fold rotational symmetry.}
\label{fig:layout}
\end{figure*}

\begin{figure*}[!htb]
\centering
\includegraphics[width=120mm]{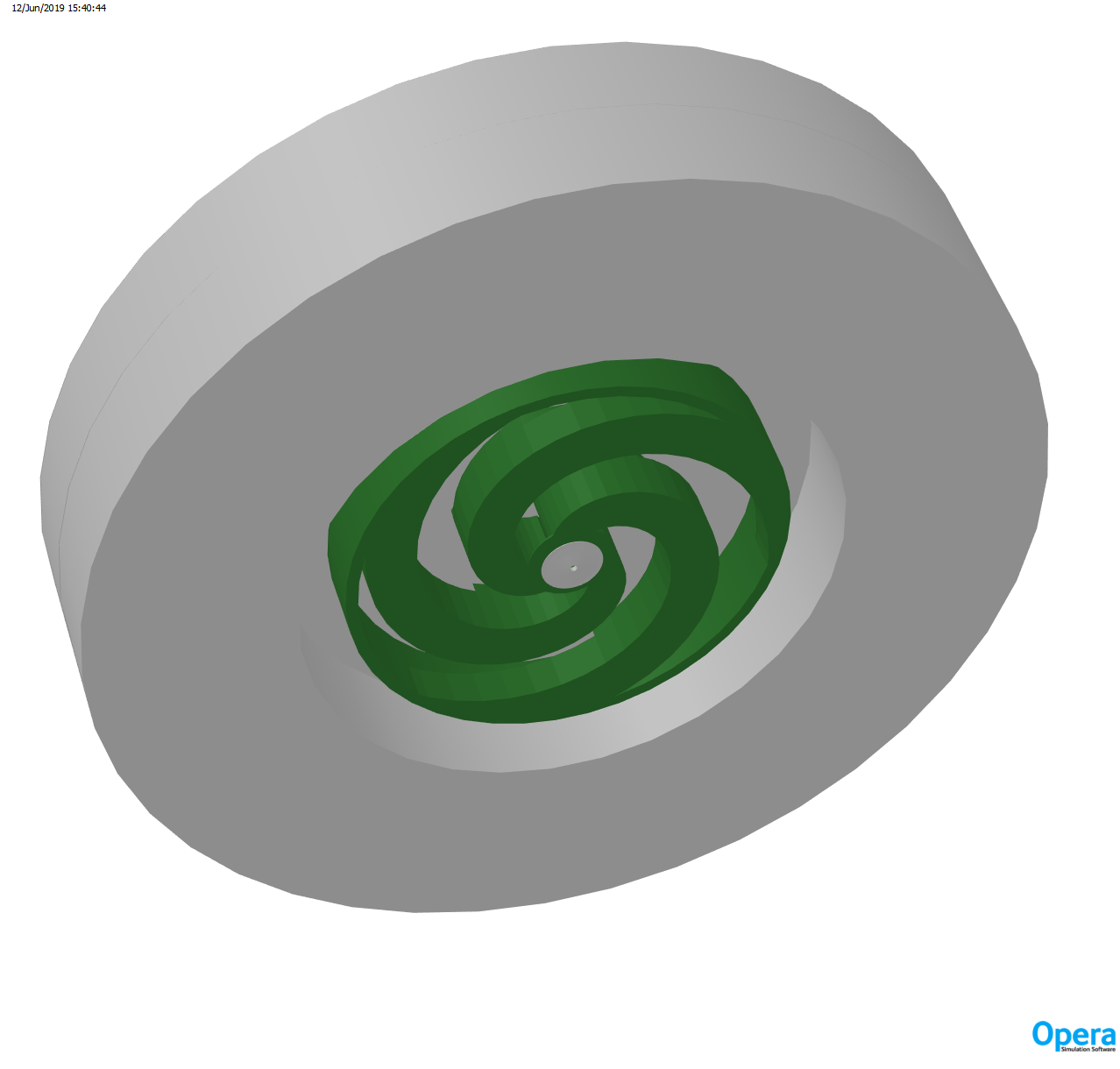}
\caption[]{Three-dimensional \textsc{opera} render of the upper half of the cyclotron magnet. The outer steel yoke (grey, diameter \SI{1340}{\mm}) has contained within it the holmium flying pole (green, diameter \SI{620}{\mm}), which is flat on the beam-facing side but to which back-cuts have been made. The weak-focusing steel cone (grey, diameter \SI{88}{\mm}) and holmium ring (green, diameter \SI{140}{\mm}) may be seen overlying the centre of the holmium pole piece. The dees, coil, and cryostat materials are not shown in this render.}
\label{fig:prxpolemesh}
\end{figure*}

The beam-facing sides of the holmium flying pole tips lie \SI{26}{\mm} from the median plane, giving a pole gap at larger radii of \SI{52}{\mm} that is sufficient for the pole cryostat walls, dees and circulating beam. In the central region there is an additional steel cone (with aperture for external injection) and accompanying holmium ring attached to the main pole to produce a negative field gradient that gives weak focusing at injection. Whilst the holmium poles sit within their cryostat, the cone may be situated at room temperature within the pole gap; at zero radius the beam gap between the two cones is \SI{24}{\mm}. A cross-sectional schematic of the magnet is shown in Figure~\ref{fig:layout}. A visualisation of the \textsc{opera} model is shown in Figure~\ref{fig:prxpolemesh}.

Superconducting coils are used to provide the strongest possible average field to minimise the cyclotron size, but we must also provide sufficient flutter to achieve stable AVF (strong) focusing at all beam radii. The field could be shaped by adjusting the pole thickness with radius on the beam-facing side, but this would increase the average gap between hills and hence reduce the variation in the AVF required for axial focusing; instead we apply \textit{back-cuts} on the opposite of the pole -- this is straightforward because the poles are already not in physical contact with the yoke (they are of course separated from the iron yoke by the cryostat walls). Using back-cut poles is a method that obtains the necessary isochronous field profile at all proton energies whilst maximising the AVF.

\begin{figure}[!htb]
\centering
\includegraphics[width=\columnwidth]{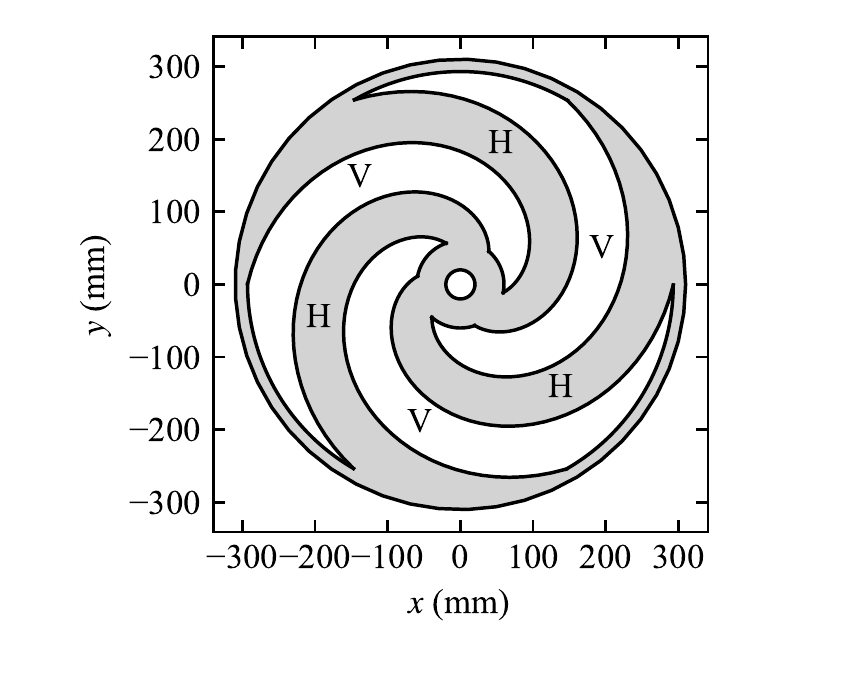}
\caption[]{Outline of the median-plane-facing side of the lower holmium pole; back-cuts are applied to the opposite side of this pole. Hills (H) and valleys (V) are shown. In the central region the hills are connected by a ring for weak focusing. At the outer radius the hills are connected by a `skirt' which helps shape the isochronous field. The central ring and outer skirt are convenient since each holmium pole may be machined and installed as a single piece.}
\label{fig:poleoutline}
\end{figure}

The pole shape is shown in Figure~\ref{fig:poleoutline}. Each holmium pole is formed into spiral hills of maximum thickness \SI{105}{\mm}; the shape is Archimedean ($r=a\theta$) with spiral parameter $a=$~\SI{70}{\mm\per\radian}. The three hills are mechanically connected by an outer `skirt' ring that shapes the isochronous field at the extraction radius, whilst an inner \SI{3}{\mm} thick holmium disk covers the poles on the beam-facing side and assists the steel cone to achieve weak focusing in the central region.

\subsection{Holmium Pole Design}
\subsubsection{Focal Requirements}
In designing an isochronous magnetic field capable of supporting charged particle orbits that are both radially- and axially-stable, one must consider the betatron motion transverse to the particle trajectory that is given by
\begin{subequations}
	\begin{equation}
		\ddot x = x_0\sin\nu_x\omega_0t\mathrm{,}
		\label{seq:x_motion}
	\end{equation}
	\begin{equation}
		\ddot z = z_0\sin\nu_z\omega_0t\mathrm{,}
		\label{seq:z_motion}
	\end{equation}
\end{subequations}
where $\nu_x$ and $\nu_z$ are the radial and axial betatron tunes respectively. For transverse particle motion that is bounded and which oscillates around the reference particle trajectory, we require real-valued tunes. The general expressions for the tunes in an $N$-sector isochronous cyclotron are approximately
\begin{subequations}
	\begin{equation}
		\nu_x^2=1-n+\frac{F(r)n^2}{N^2}+\dots\mathrm{,}
		\label{seq:radial_tune}
	\end{equation}
	\begin{equation}
		\nu_z^2=n+F(r)[1+2\tan^2\xi(r)]+\frac{F(r)n^2}{N^2}+\dots\mathrm{,}
		\label{seq:axial_tune}
	\end{equation}
\end{subequations}
where $n$ is the field index
\begin{equation}
	n=-\frac{r}{B}\frac{dB}{dr}\mathrm{,}
	\label{eq:field_index}
\end{equation}
$F(r)$ is the r.m.s. flutter
\begin{equation}
F(r)=\frac{1}{2 \pi} \int\left(\frac{B(r, \theta)-B_{0}(r)}{B_{0}(r)}\right)^{2} d \theta\mathrm{,}
\end{equation}
and $\xi(r)$ is the angle that the edge of a spiral pole sector makes with a radial line drawn from the origin of the cyclotron. In the median plane of an isochronous cyclotron $B=\gamma B_0$ increases with radius, so Eq.~\ref{eq:field_index} requires $n<0$. This requirement increases $\nu_x^2$ in Eq.~\ref{seq:radial_tune} and decreases $\nu_z^2$ in Eq.~\ref{seq:axial_tune}. Since we require real-valued tunes, the sums of the right-hand-sides of Eqs.~\ref{seq:radial_tune} and \ref{seq:axial_tune} must be positive. The challenge in designing an isochronous field is to simultaneously achieve $B=\gamma B_0$ and $\nu_z^2>0$. We see by Eq.~\ref{seq:axial_tune} that to compensate for a negative field index we must create ample flutter (i.e. an AVF with a sufficiently-varying $B$-field between the hills and valleys) and an appropriate spiral angle $\xi$. A common choice of spiral is the Archimedean spiral described in polar coordinates by $r=a\theta$, where $a$ is a constant; for an Archimedean spiral $\xi(r)=\arctan (r/a)$. In the design presented here we have found that a choice of $N=3$ sectors combined with a central field $B_0=$~\SI{4.52}{\tesla} and an Archimedean spiral with $a=$~\SI{70}{\mm \per \radian} can satisfy the requirement for an isochronous field with $\nu_z^2>0$.

In the central region of an isochronous cyclotron it is usual that the AVF is not sufficient for axial focusing. Here we look to a different method of focusing -- weak focusing. We introduce a `bump' in the radial $B$-field profile giving a negative field gradient with respect to radius and therefore a positive field index $n$; this is a typical approach used in isochronous cyclotrons. By Eq.~\ref{seq:axial_tune} we ensure $\nu_z^2>0$ and hence obtain axial stability. When employing the method of weak focusing in the central region of the cyclotron it must be remembered that the introduced $B$-field `bump' results in ions that have angular frequency $\omega$ greater than the isochronous value $\omega_0$ and will therefore eventually fall out of phase with the accelerating RF voltage, leading also to longitudinal dilution. Growing phase error may be mitigated by adjusting the initial proton phase such that the protons cross the dee gaps within a small phase window over the whole acceleration cycle, where they should certainly remain within 40$^\circ$ of the peak voltage.

\subsubsection{Pole Design}
\label{ssec:design}
Back-cuts were applied to the holmium poles using the \textsc{opera} code and were adjusted manually using results derived from the cyclotron codes \textsc{genspeo} and \textsc{z3cyclone} \cite{operamanual,gordongenspeo,z3cyclonemanual}. A limited number of azimuthally-symmetric cuts (steps) were applied which will be straightforward to manufacture; similar steel poles have been machined by us for a lower-energy cyclotron using comparable numbers and sizes of cuts. A further advantage of back-cut poles is that using a small number of cuts gives lower undulations around the desired isochronous field profile. Our pole profile gives a minimum valley field of \SI{4.1}{\tesla} and a maximum hill field of \SI{5.4}{\tesla}; the overall median-plane ($z=0$) field is shown in Figure~\ref{fig:b}. 

\begin{figure}[!htb]
\centering
\includegraphics[width=\columnwidth]{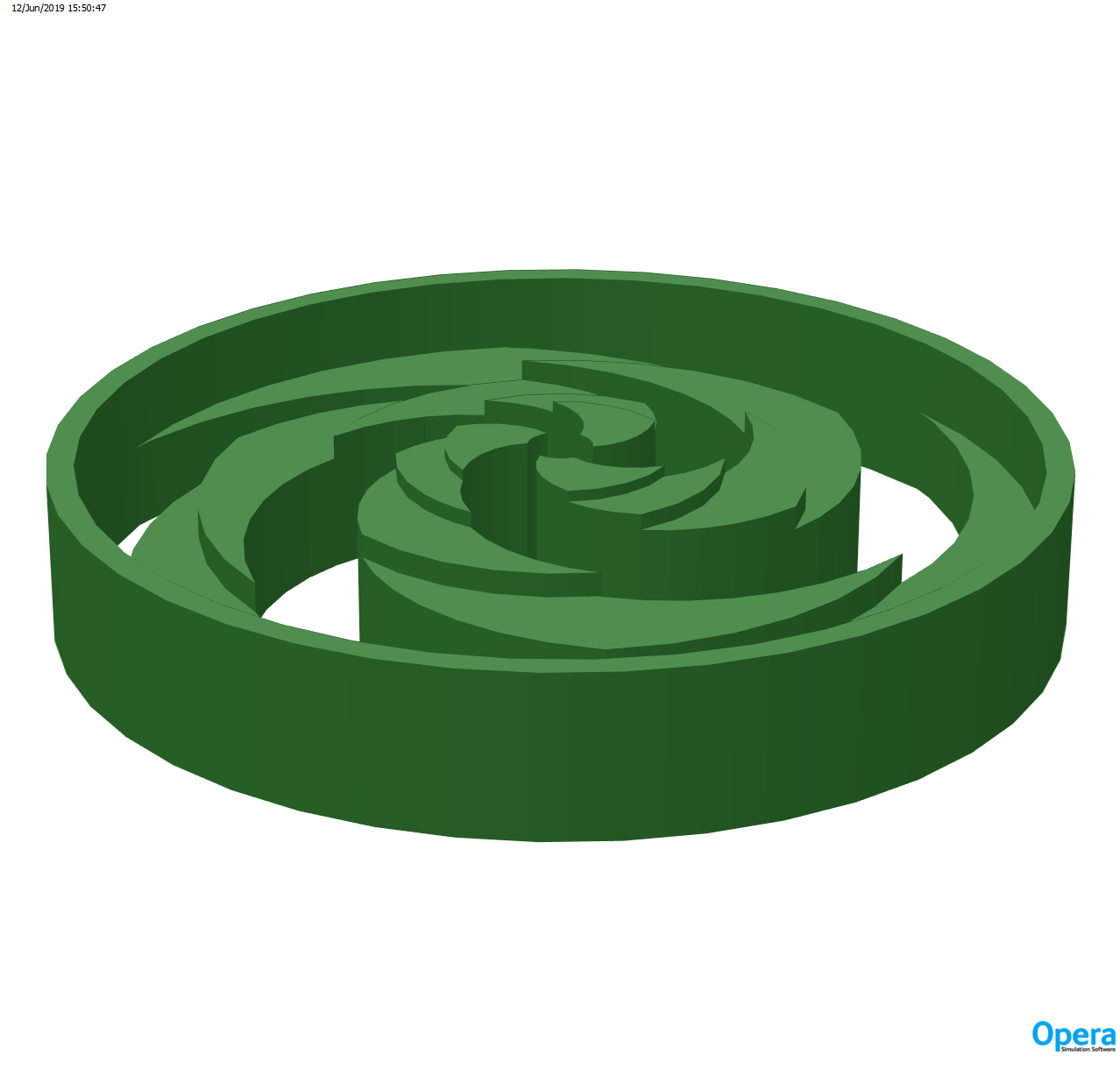}
\caption[]{Three-dimensional \textsc{opera} render of back-cut holmium pole (diameter \SI{620}{\mm}). Six cuts are sufficient to achieve the required field variation and quality.}
\label{fig:pole}
\end{figure}

\subsection{Weak-Focusing Central Region Design}
\label{ssec:weak_ring}
\begin{figure}[!htb]
\centering
\includegraphics[width=\columnwidth]{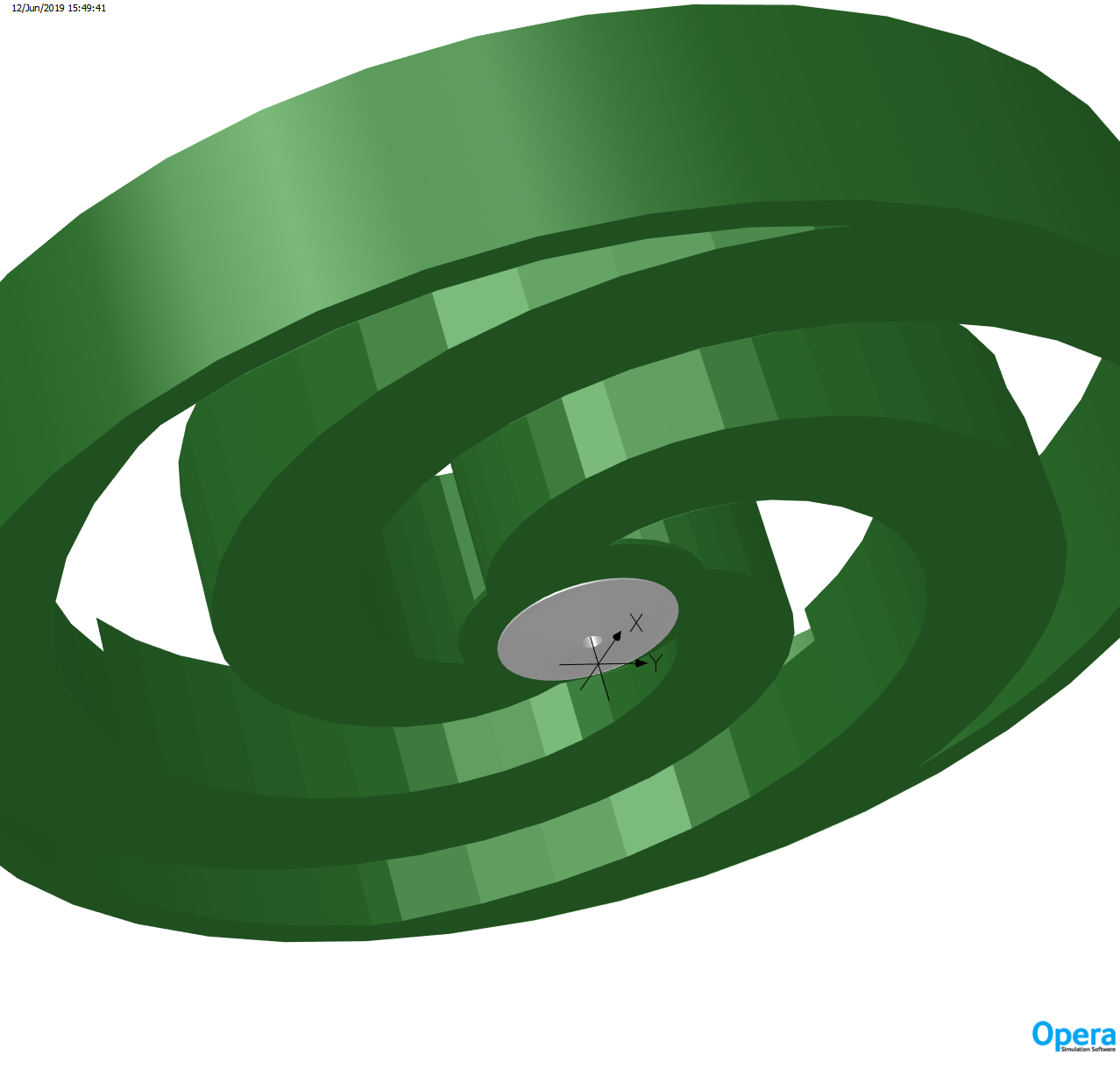}
\caption[]{Three-dimensional \textsc{opera} render of steel cone (grey, diameter \SI{88}{\mm}) and holmium ring (green, diameter \SI{140}{\mm}) lying over the holmium pole as seen from median plane.}
\label{fig:weak_focusing}
\end{figure}

No flutter is possible in the central region so weak focusing must be employed by applying a negative field gradient to maintain axial focusing; this is achieved using a steel cone and holmium ring shown in Fig.~\ref{fig:weak_focusing}. The field resulting from the introduction of the weak focusing components can be seen as a `bump' in the field at low radius in Fig.~\ref{fig:b_isoc}.

\subsection{Superconducting Coil}
\begin{figure}[!hbt]
	\includegraphics[]{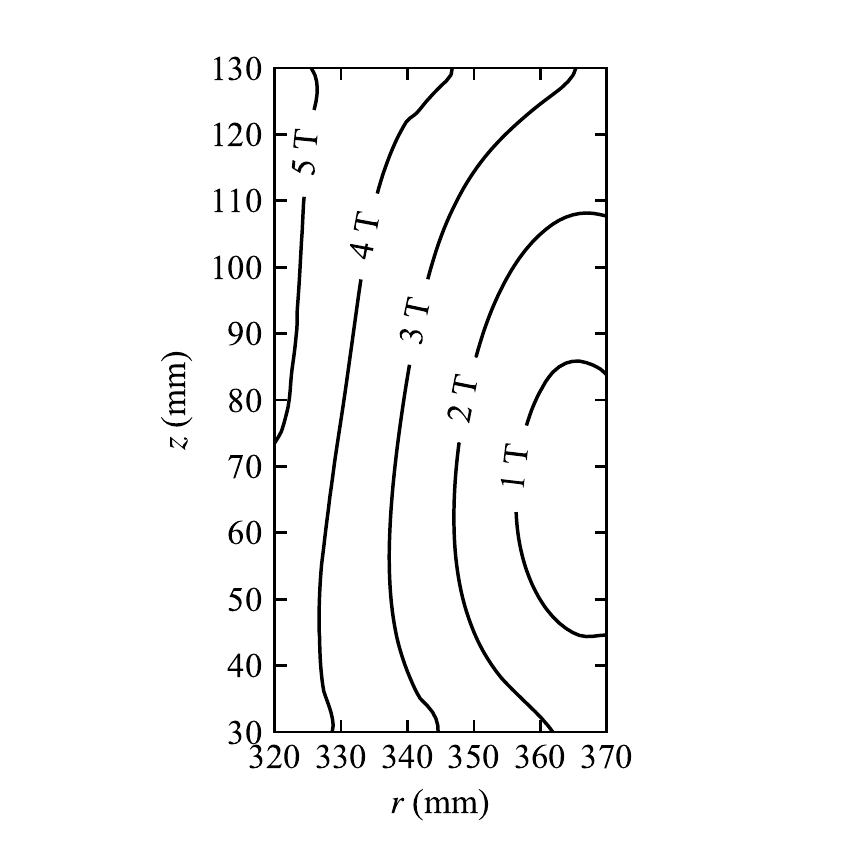}
	\caption[]{Contours showing the variation of magnetic field strength within a cross-section of the superconducting coil pack; the cross-section has been chosen through the azimuth at which the flux density is strongest. $r=$~\SI{320}{\mm} indicates the inner edge of the coil ring and $z=$~\SI{30}{\mm} indicates the beam-facing edge. Over the coil pack cross section the field remains low enough that NbTi conductors may be used, since the critical field of NbTi is typically between 8 and \SI{9}{\tesla}.}
	\label{fig:coil_field}
\end{figure}

Each superconducting coil pack sits with its beam-facing side \SI{30}{\mm} from the median plane, the poles sharing a cryostat either side of the room-temperature dees and vacuum vessel. Penetrations are made through the cryostats for dee stems, cavities, and mechanical supports. The coil parameters are shown in Table~\ref{tab:sccoilparameters}. The variation of magnetic field through the coil pack is shown in Figure~\ref{fig:coil_field}, and confirms that NbTi conductor may be reliably used.
\begin{table}[!hbt]
\caption{Superconducting coil parameters for the \SI{70}{\MeV} cyclotron.}
\centering
	\begin{ruledtabular}
        \begin{tabular}{l l}
        \bf{Parameter} & \bf{Value} \\
        \hline
        Conductor Material & NbTi\\
        Conductor Type & Cable in Channels\\
        Total Ampere-Turns & 635~At\\
        Current Density & \SI{127}{\A\per\mm\squared} \\
        Number of Turns & 1250 \\
        \end{tabular}
        \end{ruledtabular}
\label{tab:sccoilparameters}
\end{table}

\section{Median-Plane Magnetic Field Analysis}
\begin{figure}[!h]
	\includegraphics[width=\columnwidth]{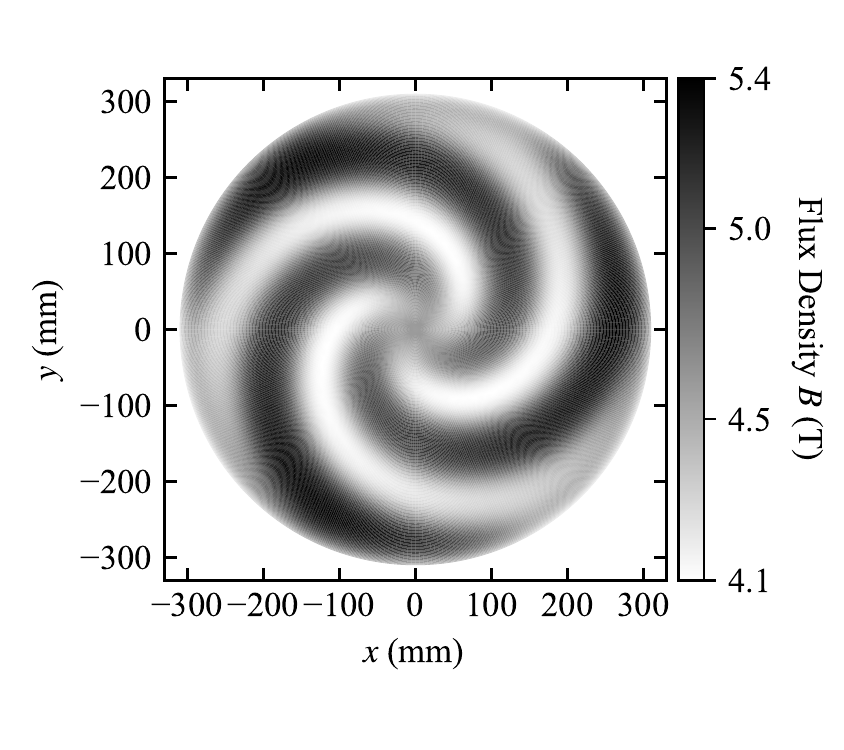}
	\caption[]{Density map indicating the strength of the vertical median-plane ($z=0$) field; the spiral focusing is visible. The minimum valley field is \SI{4.1}{\tesla} and the maximum hill field is \SI{5.4}{\tesla}, whilst the azimuthally-averaged field strength increases with radius.}
	\label{fig:b}
\end{figure}

\begin{figure}[!h]
	\includegraphics[width=\columnwidth]{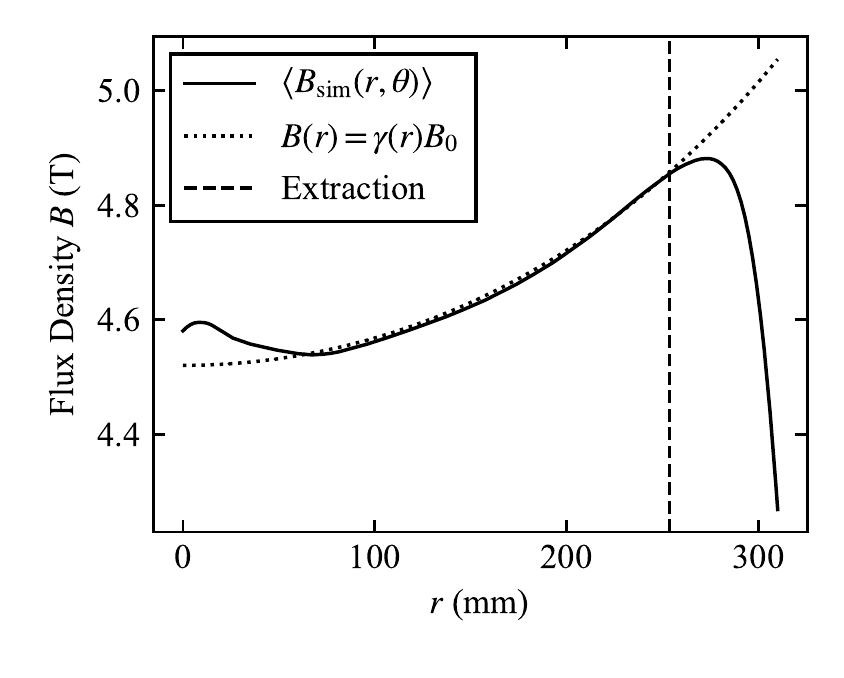}
	\caption[]{Azimuthally-averaged flux density $\langle B_\mathrm{sim}(r,\theta)\rangle$ for the \textsc{opera} field map, compared to the ideal isochronous field $B(r)=\gamma(r)B_0$.}
	\label{fig:b_isoc}
\end{figure}

We now discuss the median-plane magnetic field produced by the magnet components discussed in Section~\ref{sec:magnet_design}. The median plane at $z=0$ in Fig.~\ref{fig:layout} (where the radial and azimuthal magnetic fields fall to zero) is the plane in which particles orbit, since the magnetic flux per unit area is largest. Vertical oscillations of particles, i.e. in the $z$-direction, occur out of this plane and the magnetic field encountered by these particles was calculated by an expansion of the median-plane field. A density map of $B(r, \theta, 0)$ is shown in Fig.~\ref{fig:b}. This field profile has been achieved through carefully-chosen cuts on the side of the holmium pole that faces away from the median plane, termed `back-cuts' (described in Section~\ref{ssec:design}). Cuts have been chosen to produce a field that satisfies the criteria for orbit stability. One of these criteria is that the azimuthally-averaged radial profile of the the field shown in Fig.~\ref{fig:b} should be isochronous. Figure~\ref{fig:b_isoc} shows a comparison of the azimuthally-averaged $B$-field simulated by \textsc{opera} with the ideal value $B(r)=\gamma(r)B_0$. We see close agreement between the simulated and the ideal isochronous curves and, as we will see in Section~\ref{ssec:particle_tracking}, the agreement is sufficient for particles to stay within \SI{90}{\degree} of the phase at which the RF voltage peaks. 

We see in Fig.~\ref{fig:b_isoc} that there is a `bump' in the azimuthally-averaged field $\langle B_\mathrm{sim}(r,\theta)\rangle$ in the region $r=$~0~--~\SI{80}{\mm} which takes $B_\mathrm{sim}$ above the ideal isochronous value. This is the effect of the weak-focusing and transition rings discussed in Section~\ref{ssec:weak_ring}, which have been deliberately included to introduce a positive field index $n$ and thereby to give a region of axial stability at the cyclotron centre. This focusing mechanism is in accordance with Eq.~\ref{seq:axial_tune}; it is required in the central region of an AVF isochronous cyclotron as here the flutter term $F(1+2\tan^2\xi{r})$ would not be sufficient to cancel the negative field index $n$ that would exist if the isochronous field extended into the centre of the cyclotron. It must be remembered that an increase in the $B$-field above the isochronous value causes particles to orbit with angular frequency $\omega$ greater than the isochronous value $\omega_0$; as such, particles are prone to eventually fall out of phase with the RF and care must therefore be taken that they remain within the phase window so as not be decelerated and lost. To maximise the orbit radius range over which the weak-focusing region may exist, a phase offset is applied to particles injected into the central region of the cyclotron such that they initially lag the RF but are still accelerated; they therefore `catch up' with the RF phase before reaching the isochronous region of the $B$-field. The success of this method is demonstrated in the particle tracking simulation of Section~\ref{ssec:particle_tracking}.

\begin{figure}[!h]
	\includegraphics[width=\columnwidth]{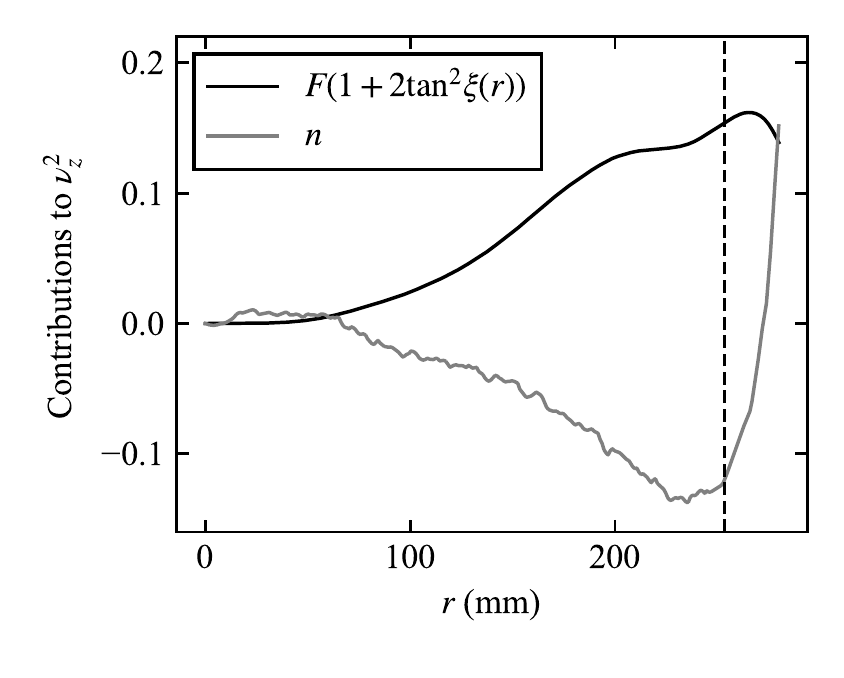}
	\caption[]{Median-plane flutter multiplied by the spiral factor term $F(1+2\tan^2\xi(r))$ compared to the field index $n$; these are the major competing terms that determine axial stability.}
	\label{fig:flutterandfield_index}
\end{figure}

Figure~\ref{fig:flutterandfield_index} shows the trade-off between the field index and flutter terms which contribute to $\nu_z^2$ in the calculated $B$-field. It is useful to look at this figure with Eq.~\ref{seq:axial_tune} in mind: we note the weak flutter term in the region $r=$~0~--~\SI{80}{\mm} which is compensated by the positive field index caused by the introduced `bump' in $B$-field. In the region $r>$~\SI{80}{\mm} we have an isochronous field profile and so the field index $n$ becomes negative; however, here the flutter term increases sufficiently to compensate. It is important to note that Eq.~\ref{seq:axial_tune} is only an approximate expression for the axial tune and that Fig.~\ref{fig:flutterandfield_index} is a comparison plot of just the first two terms -- although these terms are dominant. In practice the axial tune must be determined by numerical integration of differential equations describing the phase space of a charged particle in a magnetic field derived in Ref.~\cite{gordongenspeo}, the result of which is given next.

\subsection{Equilibrium Orbits}
\label{ssec:equilibrium_orbits}
\begin{figure}[!h]
	\includegraphics[width=\columnwidth]{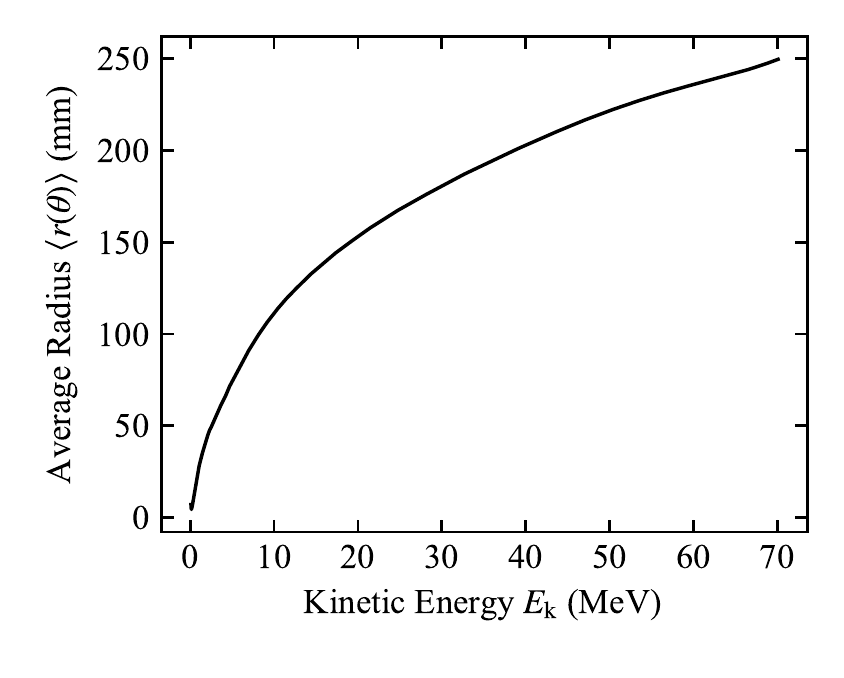}
	\caption[]{Azimuthally-averaged radii of equilibrium orbits for the 70~MeV cyclotron design, calculated from the \textsc{opera}-modelled field using \textsc{genspeo}.}
	\label{fig:radiusvenergy}
\end{figure}

Equilibrium orbits are closed orbits in a magnetic field which correspond to a particle of given mass, charge, and energy; it is conventional to design a cyclotron so that equilibrium orbits with stable radial and axial focusing exist at all energies from injection to extraction. These have been calculated here using the code \textsc{genspeo}~\cite{gordongenspeo}, which is well-validated and has been used for the design of many operating cyclotrons. Figure~\ref{fig:radiusvenergy} shows that equilibrium orbits up to \SI{70}{\MeV} exist in our field design.

\begin{figure}[!h]
	\includegraphics[width=\columnwidth]{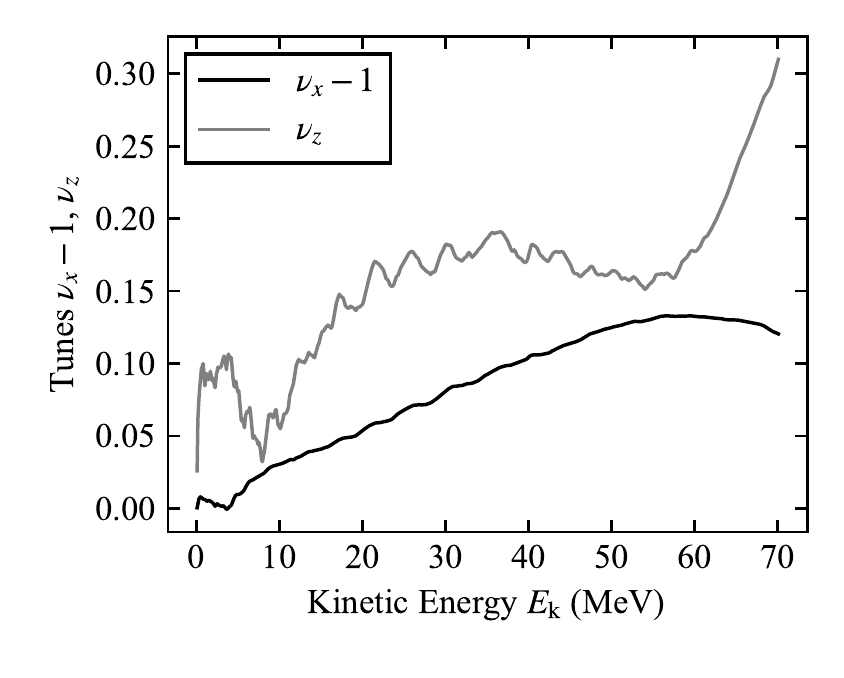}
	\caption[]{Variation of radial tune $\nu_x$ and axial tune $\nu_z$ with energy in the 70~MeV cyclotron design, calculated from the \textsc{opera}-modelled field using \textsc{genspeo}.}
	\label{fig:tunesvenergy}
\end{figure}

 \textsc{genspeo} can also calculate the radial and axial tunes $\nu_x$ and $\nu_z$ in our $B$-field. For oscillatory solutions to Eqs. \ref{seq:x_motion} and \ref{seq:z_motion} we require that $\nu_x$ and $\nu_z$ are real; Fig.~\ref{fig:tunesvenergy} confirms that the tunes are real over the full acceleration range. We note a dip in $\nu_z$ between kinetic energies of \SI{5}{\MeV} and \SI{12}{\MeV}. This occurs in the transition region from weak to AVF focusing and will result in an increase in the amplitude of axial oscillation of a particle (see Fig.~\ref{fig:zvturnnumber}, discussed later). As long as the axial oscillation amplitude remains within the bounds set by the dee aperture, the growth is tolerable; nevertheless some further field refinement may be possible to mitigate this tune dip so as to increase the tolerance of the cyclotron to manufacturing errors.

\begin{figure}[!h]
	\includegraphics[width=\columnwidth]{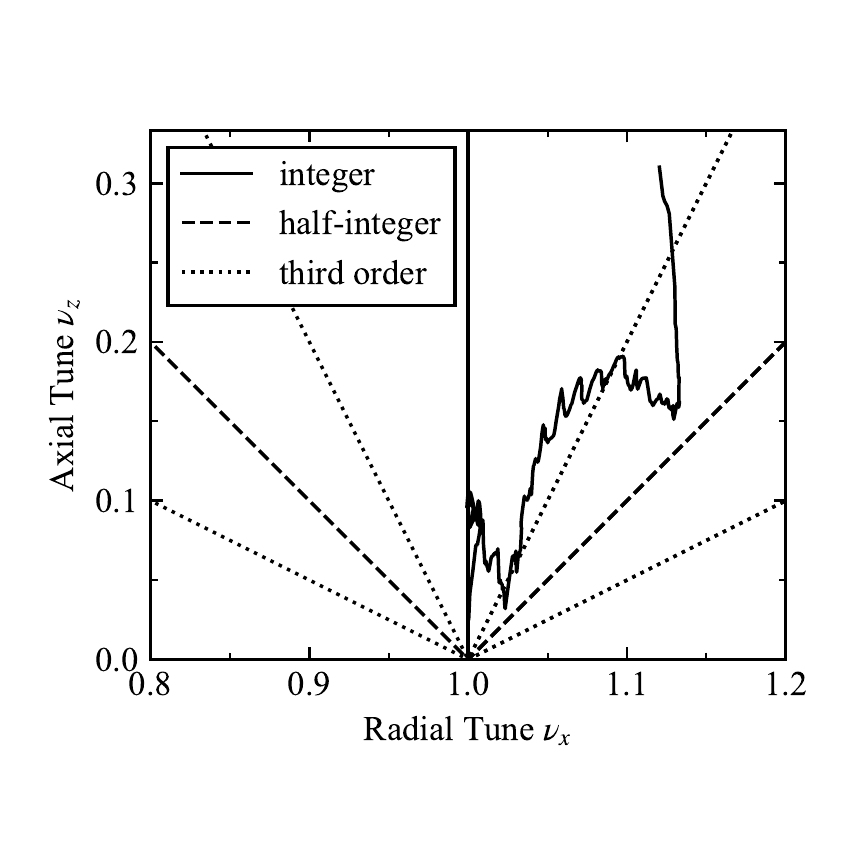}
	\caption[]{Variation of radial and axial tunes with energy in the \SI{70}{\MeV} cyclotron design, calculated from the \textsc{opera}-modelled field using \textsc{genspeo} and shown as a working point diagram. Resonances up to third order are shown.}
	\label{fig:resonances}
\end{figure}

Fig.~\ref{fig:resonances} shows the working point diagram for the \SI{70}{\MeV} design. There are two resonances which $(\nu_x,\nu_z)$ does not cross quickly during acceleration (i.e. which take place over several turns). The first of these is the $\nu_x=1$ resonance which results in a growth in radial amplitude in the central region of the cyclotron. The second is the $2\nu_x-\nu_z=2$ resonance which is crossed three times. The first crossing causes an increase in axial amplitude at around turn 50 in the cyclotron (see Fig.~\ref{fig:zvturnnumber}). The second crossing occurs around turn 170, but particle tracking performed with \textsc{z3cyclone} (see below) shows that this does not cause significant increase in axial amplitude (Fig.~\ref{fig:zvturnnumber}). The third and final crossing is very fast and is unlikely to cause axial amplitude growth.

\subsection{Particle Tracking}
\label{ssec:particle_tracking}
\begin{figure}[!hbt]
	\includegraphics[width=\columnwidth]{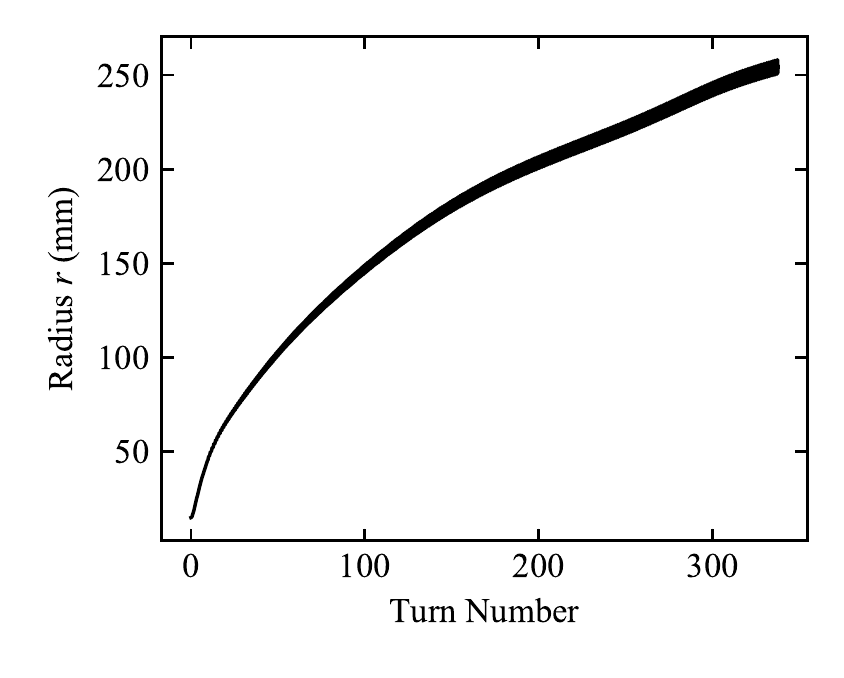}
	\caption[]{Variation of proton radius with turn number, obtained using \textsc{z3cyclone}; the apparent thickness of the radius is due to the variation of radius with azimuth within a single turn -- the orbit has a somewhat triangular shape at a given energy that reflects the varying bend radius from the three-fold-symmetric hills and valleys.}
	\label{fig:radiusvturnnumber}
\end{figure}

\begin{figure}[!hbt]
	\includegraphics[width=\columnwidth]{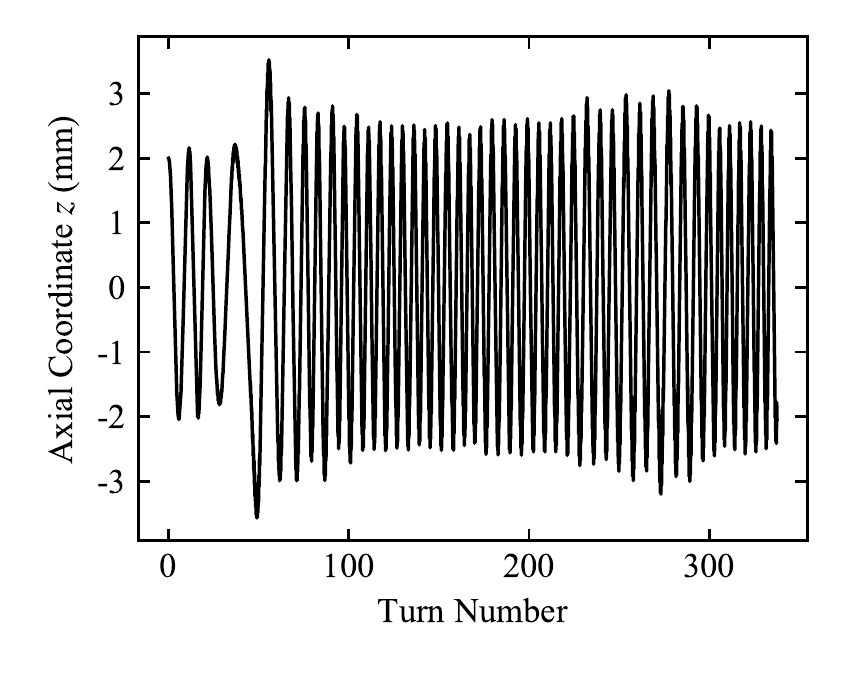}
	\caption[]{Variation of axial oscillation amplitude of a reference proton with turn number, for an example initial axial amplitude of +\SI{2}{\mm}; obtained using \textsc{z3cyclone} using the field obtained from \textsc{opera}, assuming an average voltage gain at the dee crossing of 35.7~kV. The initial axial amplitude shown is a typical value obtained in working cyclotrons, and is due to component misalignments; the overal axial amplitude growth is manageable within the available dee gap.}
	\label{fig:zvturnnumber}
\end{figure}

Particle tracking of a single proton has been performed using the code \textsc{z3cyclone} \cite{z3cyclonemanual}. This is a three-part code, with parts one and two concerning the central region of the cyclotron and part three tracking the particle to extraction energy. We have used part three to track proton amplitude through the acceleration cycle; here \textsc{z3cyclone} does not require an electric field map to describe the dee gap, and an impulse approximation for the proton energy gain across this gap (that takes into account the RF phase and transit time factor) is sufficient. Figs.~\ref{fig:radiusvturnnumber} and \ref{fig:zvturnnumber} show the radial and axial coordinates of a proton tracked through the median-plane field of Fig.~\ref{fig:b}.

\begin{figure}[!hbt]
	\includegraphics[width=\columnwidth]{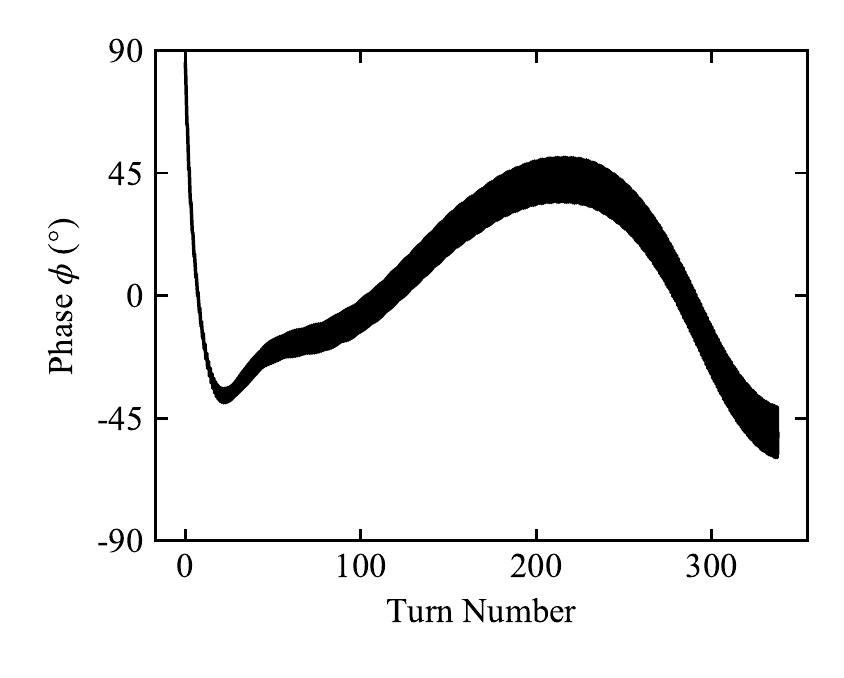}
	\caption[]{Variation of relative proton and RF phase with turn number, obtained using \textsc{z3cyclone}. Positive values of $\phi$ correspond to the RF phase leading the proton phase. This is a preliminary optimisation and may be significantly refined during magnet optimization.}
	\label{fig:phasevturnnumber}
\end{figure}

\begin{figure}[!hbt]
	\includegraphics[width=\columnwidth]{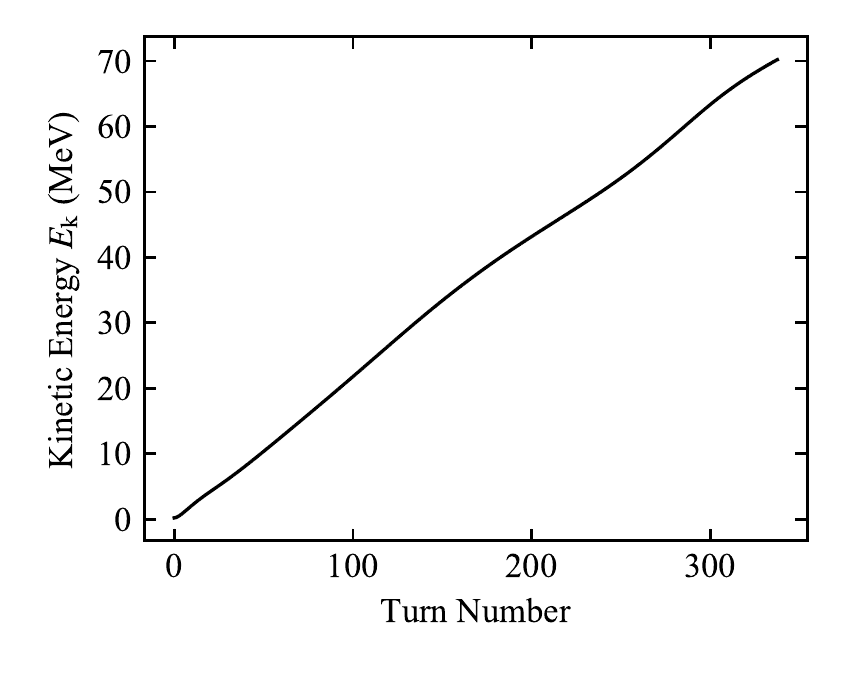}
	\caption[]{Variation of proton kinetic energy with turn number, obtained using \textsc{z3cyclone}.}
	\label{fig:energyvturnnumber}
\end{figure}

Fig.~\ref{fig:phasevturnnumber} shows the phase $\phi$ by which the RF leads or lags a reference proton. The accelerating voltage seen by a particle crossing a dee gap is given by $V=V_0\cos\phi$, where $V_0$ is the maximum voltage of the dee (\SI{40}{\kV} in this case). As long as $|\phi|<$~\SI{90}{\degree} the particle will see a positive impulse of energy. The initial phase offset is set to \SI{85}{\degree} to account for the central weak-focusing region over which the orbital frequency of the particle is greater than the RF frequency; in general an integral phase error of zero over the full acceleration cyclotron is required, and as can be see in Fig.~\ref{fig:phasevturnnumber} that has been reasonably achieved. Our design requires 336 turns for the proton to reach \SI{70}{\MeV}, with an average energy gain of \SI{34.7}{\kV} per gap crossing; this is a reasonable number of turns for a 70~MeV proton cyclotron. Fig.~\ref{fig:energyvturnnumber} shows proton energy as a function of turn number over an acceleration cycle. We see that the rate of gain of the energy decreases in those regions where the relative phase $\phi$ is furthest from \SI{0}{\degree}. 


\section{Summary and Outlook}
Combining the use of a rare-earth holmium pole with a room-temperature yoke and superconducting coil is a new method for magnet design that may be termed hyperferric, in comparison to conventional superferric magnets that use ordinary iron pole tips. Hyperferric magnets allow greater flux concentration and thereby open a route to greater focusing variation than hitherto possible. We have demonstrated a design of a superconducting cyclotron that makes use of this advantage, and which is the first realistic isochronous \SI{70}{\MeV} proton cyclotron with an average field above \SI{4}{\tesla}; the drastic reduction in cyclotron size to a yoke mass of less than 9~tonnes -- around one order of magnitude smaller than existing approaches -- makes it very attractive for a variety of uses such as particle therapy and isotope production, and there are no particular barriers to delivering a high dose rate suitable for such emerging techniques as FLASH radiotherapy. Cyclotrons are the workhorse proton source across many areas of industry, medicine and physical research, and our approach greatly increases the accessibility of such sources to users in a variety of disciplines.

We have also demonstrated the advantages of a flying pole design, which at the same time allows the use of both a small cold mass and a back-cut pole; the latter maximises the beam-plane flux density whilst providing sufficient AVF focusing, both important in minimising cost. The resulting cyclotron -- with a diameter of 1340~mm and height of 862~mm -- is far smaller than any other source of monochromatic, high-current protons yet proposed, and far smaller than existing normal-conducting cyclotrons of the same energy. For example, such a source may be used for ocular therapy within the typical room size of an IMRT system, and our approach may be scaled up within limits to deliver protons of higher kinetic energy for other purposes.

The hyperferric approach is not only useful for cyclotron design. The general approach of a cold, shaped flying holmium pole should be of interest in other types of magnetic system, and we envisage it may be utilised in such systems as high-gradient quadrupoles for particle accelerators, high-field wigglers for synchrotron radiation production, and applications where a compact, planar field of several tesla is desired; our back-cut field shaping method may be of benefit also in several of those applications.

\begin{acknowledgments}
We gratefully acknowledge support for this work from the Antaya Foundation for Science and Technology, and from the Science and Technology Facilities Council under Grant Nos. ST/G008248/1 and ST/R002142/1.
\end{acknowledgments}


\providecommand{\noopsort}[1]{}\providecommand{\singleletter}[1]{#1}%

\end{document}